\documentclass[%
 reprint,
 amsmath,amssymb,
 aps,
floatfix,
]{revtex4-2}

\usepackage{graphicx}
\usepackage{dcolumn}
\usepackage{bm}
\usepackage{float}

\begin{document}

\title{Structure-based approach can identify driver nodes in ensembles of biologically-inspired Boolean networks}

\author{Eli Newby}
\email[]{eyn5@psu.edu}
\affiliation{Department of Physics, The Pennsylvania State University, University Park, PA 16802}
\author{Jorge G\'{o}mez Tejeda Za\~{n}udo}
\affiliation{Eli and Edythe L. Broad Institute of MIT and Harvard, Cambridge, MA 02142}
\affiliation{Department of Medical Oncology, Dana-Farber Cancer Institute, Harvard Medical School, Boston, MA 02115}
\author{R\'{e}ka Albert}
\affiliation{Department of Physics, The Pennsylvania State University, University Park, PA 16802}
\affiliation{Department of Biology, The Pennsylvania State University, University Park, PA 16802}

\date{\today}

\begin{abstract}

Because the attractors of biological networks reflect stable behaviors (e.g., cell phenotypes), identifying control interventions that can drive a system towards its attractors (attractor control) is of particular relevance when controlling biological systems. Driving a network's feedback vertex set (FVS) by node-state override into a state consistent with a target attractor is proven to force every system trajectory to the target attractor, but in biological networks, the FVS is typically larger than can be realistically manipulated. External control of a subset of a biological network's FVS was proposed as a strategy to drive the network to its attractors utilizing fewer interventions; however, the effectiveness of this strategy was only demonstrated on a small set of Boolean models of biological networks. Here, we extend this analysis to ensembles of biologically-inspired Boolean networks. On these models, we use three structural metrics — PRINCE propagation, modified PRINCE propagation, and CheiRank — to rank FVS subsets by their predicted attractor control strength. We validate the accuracy of these rankings using three dynamical measures: \textit{To Control, Away Control}, and logical domain of influence. We also calculate the propagation metrics on effective graphs, which incorporate each Boolean model's functional information into edge weights.
While this additional information increases the predicting power of structural metrics, we find that the increase with respect to the unweighted network is limited. The propagation metrics in conjunction with the FVS can be used to identify realizable driver node sets by emulating the dynamics that are prevalent in biological networks. This approach only uses the network's structure, and the driver sets are shown to be robust to the specific dynamical model.
\end{abstract}

\maketitle


\section{Introduction}

In the modeling of complex systems, networks are a powerful tool for capturing the interplay between many interacting components. A network $\mathcal{G}(\mathcal{N},\mathcal{E})$ is a set of nodes $\mathcal{N}$ of size $N$ that represent the components of a system and are connected by a set of edges $\mathcal{E}$ that represent the interactions between components. Once built, networks can illuminate emergent properties that are not visible at the individual component level \cite{Newman2003-SIAM,Watts2004-ARSoc,Milo2002-Sci}. For example, in cell biology, networks are created to model interactions among (macro)molecules and the totality of these interactions determine cellular behaviors. In these models, the nodes represent biological macromolecules such as genes or proteins and the edges represent different types of interactions such as transcription, chemical reactions, or regulation \cite{Alm2003-StructBio,Zhu2007-GD}.

A dynamic model appended to the network structure describes the time evolution of the biological system \cite{Abou-Jaoude2016-FrontGenet,Wynn2012-IntegBio,G_T_Zanudo2018-SystBio,Alon2006-CRC,Tyson2003-CellBio}. In the discrete time case, which we will focus on in this manuscript, each node's state is represented as $x_i(t+1) = F_i(X_{I_i}(t))$ where $x_i(t)$ represents the state of node $i$ at time $t$, and $F_i$ represents its update function. $F_i$ depends on the vector $X_{I_i}(t)$ consisting of the states of the nodes in the set $I_i$, which is the set of nodes with edges incident on node $i$ in the network. The network at any time can be fully described by the state vector $X(t) = \{x_1(t),\ldots,x_N(t)\}$. Here, we focus on Boolean discrete dynamical models, in which each node only has two possible states $X(t) \in \{0, 1\}^N$. We use the well-established stochastic asynchronous updating scheme, where at each time step, a randomly selected node is updated according to its update function. This update scheme thoroughly probes the state space, omits unstable oscillations that depend on node synchrony, and preserves the long-time stationary states of the system \cite{Saadatpour2010-JTheorBio}.

If we allow the model to naturally evolve in time, the system will eventually enter and be trapped in a region of the state space, called an attractor \cite{Hopfensitz2013-computStat,Klemm2005-PRE}. Attractors can be classified as point attractors (steady states) or complex attractors made up of multiple states. In cell-level biological networks, the attractors represent the cell's different phenotypes, which can be either desirable (e.g., healthy states) or undesirable (e.g., cancerous states). Studying biological network models' attractor repertoires constitutes an in silico approach to understanding how phenotypes arise. Exploring the structure of the state space reveals control interventions that alter the trajectories within the state space, which prevent undesirable phenotypes or elicit desirable phenotypes, thus driving the network to or away from one of its attractors. In the area of network control, the domain of research focused on driving complex systems to their attractors is referred to as attractor control \cite{Liu2016-ModPhys,Liu2011-Nat,Ruths2014-Sci,Cornelius2013-NatCom,Wang2016-NatCom,Murrugarra2016-BMCSystBio,Mochizuki2013-JTheorBio,Zanudo2017-PNAS}. Attractor control is particularly well-suited to study biological systems because of the relationship between attractors and biological cell states. Attractor control poses three questions: what is the best choice of nodes to drive a network to a target attractor? How can we manipulate the state of these nodes to guarantee convergence to a target attractor? And how do we implement this control action? We will focus on the first question here. One answer to the second question is the theory of feedback vertex set (FVS) control established by Mochizuki et al. \cite{Mochizuki2013-JTheorBio,Fiedler2013-DynamDiffEQ,Zanudo2017-PNAS}, which informs our search for driver nodes. For the third question, we assume that control by node-state override is possible, which has been shown for many cellular biological systems \cite{Hall2009-CellBio,Alberts2002-Garland,Gomez_Tejeda_Zanudo2019-PhysBio,Guinn2019-NucAcid}.

Attractor control through FVS control guarantees the convergence of a system into any of its attractors. A FVS of a network is a set of nodes that when removed makes the network acyclic. Here, when referring to a FVS, we consider minimal FVSs, i.e., the minimal number of nodes necessary to render the network acyclic. The minimal FVSs are not necessarily unique. It was proven that each FVS is a set of determining nodes of the dynamics of a regulatory network \cite{Fiedler2013-DynamDiffEQ,Mochizuki2013-JTheorBio}. That is, if we drive every FVS node into its state in a corresponding attractor, then the system will converge to that attractor. Mochizuki et al. originally formulated this approach on systems of nonlinear ODEs, but here we apply it to systems of Boolean update functions, which is a natural extension of the approach. For more on the formalization of the FVS and its extension to Boolean systems see Appendix \ref{append:FVS}. 

While FVS control answers the first two questions posed by attractor control, here, to enhance applicability, we investigate how to reduce the number of driver nodes. The size of a minimal FVS in biological networks is typically too large to realistically drive every required node. Previously, we have shown for Boolean models of biological networks that three propagation metrics — PRINCE Propagation, modified PRINCE Propagation, and CheiRank — can accurately rank FVS subsets by their ability to drive a network \cite{Newby2022-Chaos}, so driving the full FVS is not necessary for attractor control. While these results were consistent for multiple biological models, they were not an exhaustive analysis of the relationships between FVSs, propagation metrics, and the systems' dynamical models. Here we use multiple ensembles of Boolean models with biologically-inspired network structures to comprehensively probe these relationships and the manner in which the structure of networks influences their controllability, without requiring the creation and validation of many real Boolean biological models.

\section{Results}
\subsection{Biologically-inspired networks and dynamic models}
\subsubsection{Networks were randomly generated with topologies that emulate biological networks}

To create a test ensemble of models, we generated networks that reflect the structural properties of biological networks and are random otherwise. A meta-analysis of 132 Boolean models of biological networks performed by Kadelka et al. found that these networks have a Poisson in-degree distribution and a power law out-degree distribution with an average in-/out-degree of 2.37 \cite{Kadelka2020-arXiv}. Utilizing the configuration model \cite{Newman2018-OxUP}, we generate networks of 50 nodes whose edges are distributed according to degree distributions found by Kadelka et al. We add out-going stubs to each node following a power law out-degree distribution and connect the stubs to nodes uniformly at random. Because the outgoing connections are chosen uniformly at random, the in-degree distribution will follow a Poisson distribution \cite{Gui2001-Numer}. Kadelka et al. also found that on average 75.1\% of edges are activating, so we randomly specified with probability 0.75 that the created edge was positive; it was negative otherwise.

\begin{figure}[ht]
    \centering
    \includegraphics[width=\linewidth]{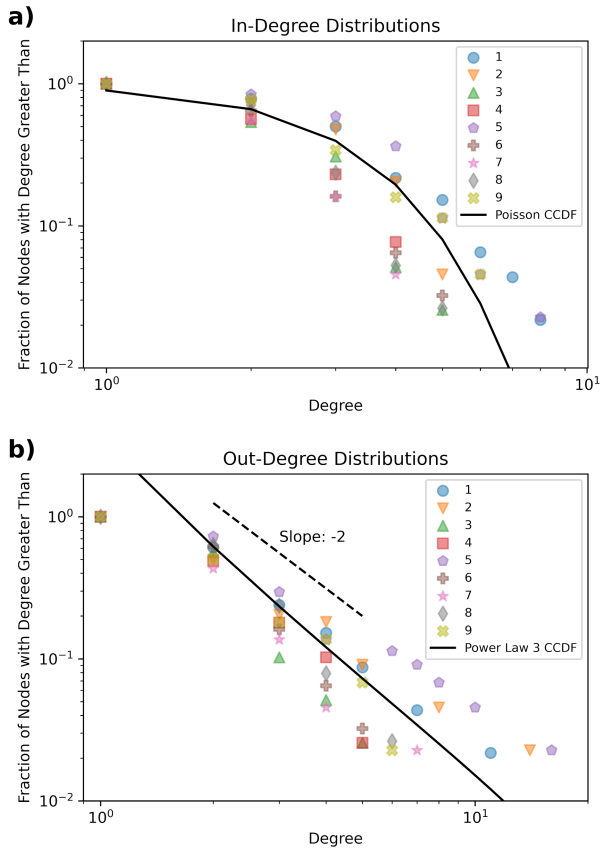}
    \caption{Complementary cumulative distribution functions (CCDFs) of in- and out-degree distributions of the nine generated networks after pruning. On these plots, the value on the y-axis indicates the probability that the degree is greater than or equal to the degree on the x-axis. The in-degree distributions along with the Poisson's CCDF (a) and the out-degree distributions along with the power law's CCDF (b) show that the networks still have the correct degree distributions after pruning. Panel b includes a guide to the eye indicating slope 2 to show that the complementary cumulative out-degree distributions follow a power law with power 2, which shows that the out-degree distributions follow a power law with power 3.
}
    \label{fig:fig1}
\end{figure}

After creating random networks, we pruned any repeated edges and reduced each network to its largest strongly connected component. We removed the source nodes of each network because we wanted to focus on the effect of FVSs on the attractor repertoire, not on the selection of specific attractors in response to source nodes. Outside of the source nodes, the largest SCC is the largest contributor to multi-stability because it contains most nodes of the minimal FVS, so we focus our analysis on it by removing any out-components. If a network is composed of many large SCCs, our analysis can easily be adapted by treating each large SCC as an individual network and then aggregating the results. Only networks that consist of at least 30 nodes after reduction were kept. We confirmed that the reduction to the strongly connected component did not strongly affect the degree distribution of the networks. Over the nine remaining network topologies the average in-/out-degree is 2.27, with average degrees ranging from 1.75 to 3.02. Figure \ref{fig:fig1} shows the complementary cumulative in- and out-degree distributions for the nine networks. After reduction, the in-degree distributions still follow a Poisson distribution (Fig. \ref{fig:fig1}a),  and the out-degree distributions still follow a power law distribution (Fig. \ref{fig:fig1}b), so the pruned networks still reflect the properties found by Kadelka et al.

These reduced networks have FVSs of sizes ranging from 5 to 9. Because identifying minimal FVSs is an NP-hard problem, we follow Za{\~n}udo et al. and identify FVSs using a simulated annealing algorithm that efficiently identifies near-minimal FVSs \cite{Zanudo2017-PNAS,Galinier2013-Heur}. When determining FVS subsets, we consider every subset of any identified FVS. For example, each topology has between 9 and 20 candidate one-node FVS subsets, which indicates that there are multiple near-minimal FVSs for every network. We label the nine networks according to the number of candidate one-node FVS subsets from greatest to smallest with ties arbitrarily broken, so network \#1 is the network with the largest number of one-node FVS subsets. Going forward, we will refer to the set of all candidate FVS subsets as just the set of FVS subsets.

\subsubsection{Random nested canalyzing functions were used to populate the networks with ensembles of biologically-inspired Boolean models}

For each network we generate an ensemble of 100 dynamical Boolean models consistent with the topology. In Boolean models of biological networks a vast majority of the Boolean rules are nested canalyzing functions \cite{Kadelka2020-arXiv,Subbaroyan2022-PNASNexus,Murrugarra2011-JTheorBio,Kauffman1993-OxUP,Hinkelmann2012-IntScholar}. Nested canalyzing functions are a set of Boolean functions in which every input variable is either canalyzing or conditionally canalyzing \cite{Dimitrova2022-Automatica}. If a canalyzing input $n_i$ is fixed to a specific value $x_i = a_i$, then the function $F(X_{I_i})$ is fixed $F = b_i$. Fixing a canalyzing variable into its non-canalyzing state yields a reduced function. This reduced function can itself be canalyzing, then the canalyzing variables of this function are considered conditionally canalyzing. If every reduced function is also canalyzing, then the original function is a nested canalyzing function. In nested canalyzing functions, the variables can be labeled with their canalyzing layer, which denotes the number of reductions required before a node becomes canalyzing. Nested canalyzing functions are prevalent in biology because they create a stable system. Compared to a system built on random Boolean functions, systems built with nested canalyzing functions have fewer attractors and are more robust to perturbations, placing them in the so-called critical regime thought to be a key property of biological systems \cite{Nikolajewa2007-Biosys,Kauffman2003-PNAS,Kadelka2020-arXiv,Kauffman2004-PNAS}. Thus, nested canalyzing functions provide a smaller set of Boolean functions to choose from, which will create a network that accurately represents the dynamics and properties of published biological networks. We selected the Boolean rules for our model randomly from the set of nested canalyzing functions that were consistent with the edge signs. For the nine generated topologies the minimum number of possible nested canalyzing rule-sets was of the order of $10^{11}$, which gives a large set of possible rule-sets to analyze.

\begin{figure}[ht]
    \centering
    \includegraphics[width=\linewidth]{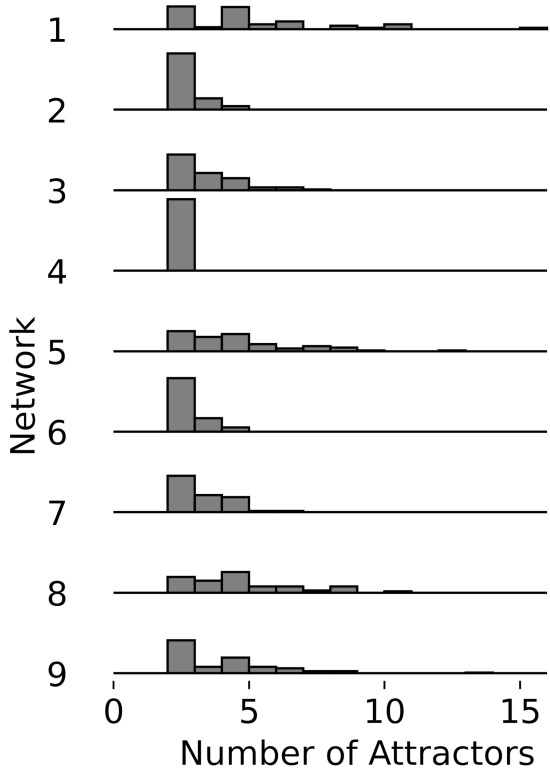}
    \caption{Histogram plots of the number of attractors of each of the 100 rule-sets for the nine networks. There are at least 2 attractors for all of the rule-sets, so it is possible to calculate attractor control values on every rule-set.}
    \label{fig:fig2}
\end{figure}

When generating our Boolean models, we only kept models that yielded more than one attractor because multi-stability is needed to test our ability to drive a model towards or away from its original attractors. We also exclude rule-sets that have an attractor wherein more than 50\% of the network oscillates because it is not possible to precisely quantify distance from such an attractor. Figure \ref{fig:fig2} indicates a histogram of the number of attractors in each topology for all 100 generated Boolean models (rule-sets). For 8 out of the 9 network topologies, we see a spread in the number of attractors across the 100 rule-sets, which is consistent with the ensembles having a diverse dynamic landscape. For one network (network \#4), all rule sets had 2 attractors, which we attribute to the large number of negative cycles in the network \cite{Thomas1981-NumericalMethods,Richard2019-JTheorBio}, but this network still has a diverse set of attractors over its 100 models. The observed diversity in the dynamics of the ensembles allows us to investigate whether the structure-based drivers we identify can control each distinct rule-set. Such finding would indicate a topological underpinning of the biological driver sets, independent of the specific dynamical model.

\subsection{Topological Propagation Metrics and Control}

Our previous study on Boolean models of real biological networks indicated that propagation metrics are a good predictor of a FVS subset's ability to drive a dynamical model \cite{Newby2022-Chaos}. In particular, the intersection of PRINCE propagation \cite{Vanunu2010-PLoS,Santolini2018-PNAS}, modified PRINCE propagation \cite{Newby2022-Chaos}, and CheiRank \cite{Zhirov2010-EPJB}, referred to as the propagation intersection metric (see Methods), accurately predicted control of real biological networks. Because we have ensembles of dynamical models for each topology, we can probe if these topologically important driver sets are successful at controlling the network regardless of the network's dynamical model. To do this, for each of the nine network topologies, we investigate the relationship between a FVS subset's propagation metrics (topology-based) and three different control metrics (dynamics-based): \textit{To Control, Away Control}, and the logical domain of influence (LDOI).

The PRINCE propagation and modified PRINCE propagation function by introducing a constant perturbation $P_{n_i}$ onto a node $n_i$ and letting this perturbation flow through the network. The flow is normalized based on the in- and/or out-degrees of the nodes and a decay variable that causes degradation at each step. The two variants differ in their normalization \cite{Newby2022-Chaos}. The normalization and degradation guarantee convergence to a steady state $X^{P_{n_i}}(t) = \Pi(t)$. For each perturbation $P_{n_i}$, we assign the value of each node $n_j$ at steady state as its propagation score $\Pi_j(t)$ in response to the perturbed node. The (modified) PRINCE metric value of a node is defined by taking the average of the magnitudes of the (modified) PRINCE scores $\overline{|\Pi(t)|}$, which summarizes each perturbation's effect on the network as a single value. The (modified) PRINCE metric value is large when the perturbed node is central to the network and can easily reach every other node. The metric values also increase when the perturbed node participates in short cycles because the values of the nodes in these cycles quickly feed back onto the perturbed node, increasing its score and thus increasing the scores propagated to the rest of the nodes in the network. Because these two properties are also signifiers of dynamically influential nodes in a network, the PRINCE and modified PRINCE are expected to identify key node sets.

The CheiRank represents the chance that a random walk originated at a specific node. The algorithm for the CheiRank is a variation on the algorithm for calculating the PageRank of a network \cite{Zhirov2010-EPJB}. The PageRank predicts the probability that a random walk will end at each node. It is calculated by finding the steady state of a Markov process with an added possibility of jumping to a random node at each step. For the CheiRank calculation the edges are traversed in reverse during the Markov process. The resulting CheiRank value of each node is thus a representation of how “source-like” each node in the network is, which helps to determine which nodes are more influential on the network.

\begin{figure*}[htpb]
    \centering
    \includegraphics[width=\textwidth]{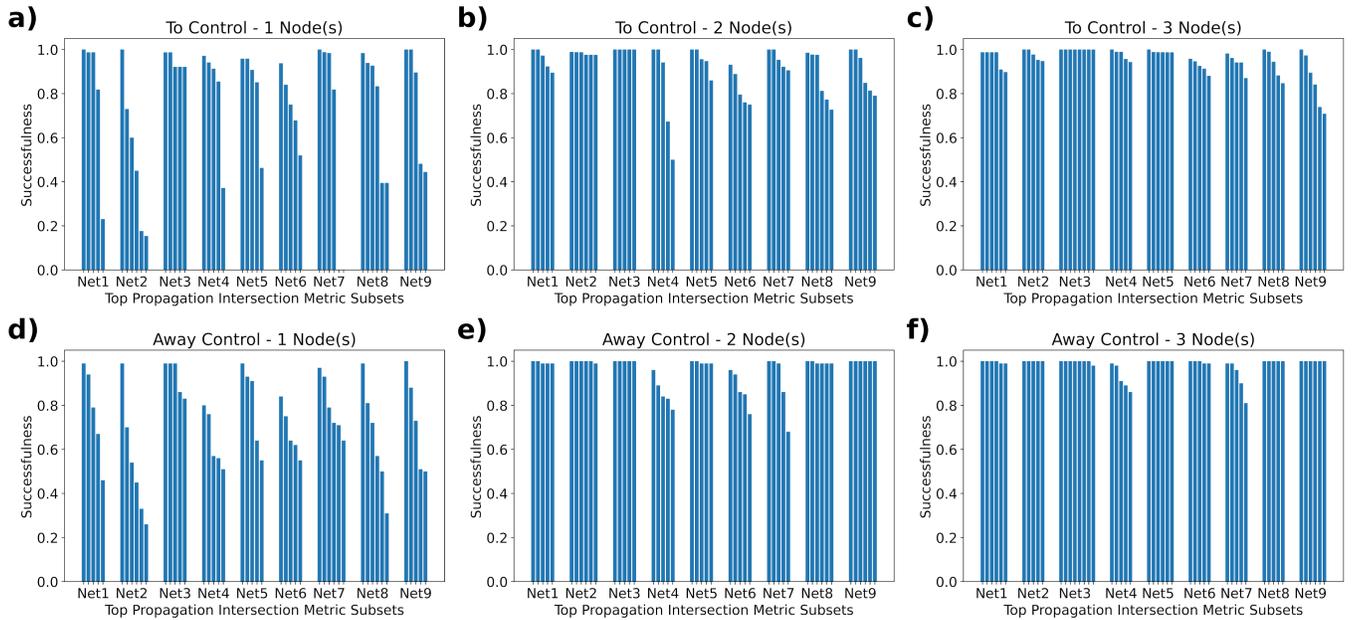}
    \caption{The \textit{To Control} successfulness (a-c) and \textit{Away Control} successfulness (d-f) values of the top 5 FVS subsets of size one (a, d), two (b, e), and three (c, f) according to the propagation intersection metric. The successfulness of a FVS subset is the percentage of rule-sets (out of 100) in which the \textit{To/Away Control} has a value greater than 0.9. When finding the top 5 FVS subsets all subsets are included in a tie. For each network, the results are ordered by their successfulnesses to aid in the comparison of the results between networks. }
    \label{fig:fig3}
\end{figure*}

For the propagation metrics to predict the dynamics of the Boolean model, they must accurately reflect how the inputs of a target node interact to determine the node's future state. The propagation metrics cannot differentiate between the ON and OFF states of a node; instead, they focus on the amount of \textit{influence} one node has over another. Taking a two-input function as an example, the propagation metrics cannot differentiate if the function is an AND or OR function, and they cannot determine whether the next state of the target node will be ON or OFF. However, the propagation metrics do recapitulate a shared property of both functions, namely that each input is equally likely to determine the state of the target node. The propagation metrics are less adept at reflecting the combinatorial nature of Boolean functions with more than two inputs. For example, in the function \\ \texttt{A *= B and (C or D)} the three inputs do not equally influence the state of \texttt{A}. In our networks, more than 65\% of the nodes have less than 3 inputs, so these nodes follow the simpler Boolean rules that can be well-estimated by the propagation metrics. Furthermore, a nested canalyzing function is better predicted by these propagation metrics than a random Boolean function because of the existence of canalyzing layers within the function. Every node within a canalyzing layer will have equal influence on the target node, so there are still some equivalences that the propagation metrics can reflect despite every input not being equivalent. For example, nodes \texttt{C} and \texttt{D} in the function \texttt{A *= B and (C or D)} are in the same canalyzing layer, and have the same influence on the state of node \texttt{A}. In summary, because the majority of the nodes in our networks have few inputs, and all the functions are nested canalyzing, we expect that the propagation metrics to accurately predict influences in the Boolean dynamics.

\subsubsection{The propagation intersection metric can distinguish FVS subsets based on their \textit{To/Away Control}}

For each topology, we use the propagation intersection metric to rank FVS subsets and pick the top five FVS subsets per network. In the case of ties, we keep all tied subsets. A FVS subset is considered successful in a singular rule-set if it achieves a \textit{To Control} or \textit{Away Control} value greater than 0.9. For each FVS subset we determine the percentage of rule-sets (out of 100) in which it is successful; we will refer to this percentage as the \textit{successfulness} of the FVS subset. Figure \ref{fig:fig3} indicates the successfulness of the top five single-node (Fig. \ref{fig:fig3}a,d), two-node (Fig. \ref{fig:fig3}b,e) and three-node (Fig. \ref{fig:fig3}c,f) FVS subsets for \textit{To Control} (Fig. \ref{fig:fig3}a-c) and \textit{Away Control} (Fig. \ref{fig:fig3}d-f). The results are ordered by decreasing successfulness to aid in comparisons. In the one-node subset case, for every network topology at least one of the top five subsets has a successfulness $\geq$ 90\%, and a majority of the identified subsets have a successfulness $\geq$ 75\%. In the two-node case, a majority of the identified FVS subsets have a successfulness $\geq$ 90\%, and in the three-node case almost every identified FVS subset has a successfulness $\geq$ 90\%. These results recapitulate the propagation intersection metric's strong predictive power of \textit{To Control} and \textit{Away Control} \cite{Newby2022-Chaos} and demonstrate that the FVS subsets ranked in the top according to the propagation metrics achieve high control over the diverse set of network dynamics.

\begin{figure}[htpb]
    \centering
    \includegraphics[width=\linewidth]{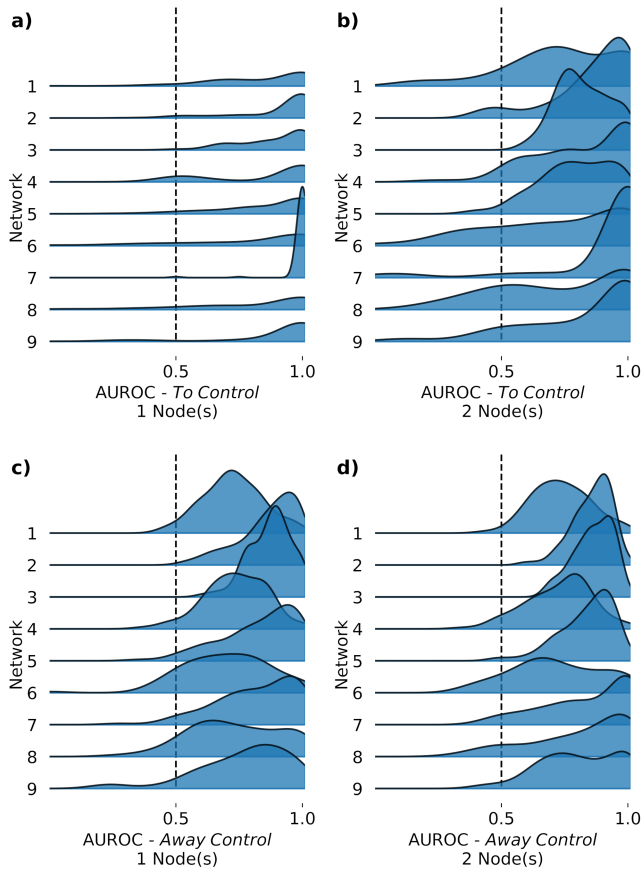}
    \caption{Distributions of individual rule-set AUROC values demonstrating the predictive ability of the propagation intersection metric. Distributions for each of the nine networks for one-node FVS subsets (a,c) and two-node FVS subsets (b,d). For each rule-set, the AUROC represents how well the propagation intersection metric classifies the binarized version of the \textit{To Control} (a,b) and \textit{Away Control} (c,d) (whose value is 1 if the \textit{To/Away Control} is greater than 0.9 and 0 otherwise). An AUROC value greater than 0.5 indicates that the propagation intersection metric performs better than randomly sorting the FVS subsets. This cutoff is indicated by the dotted lines in each plot.}
    \label{fig:fig4}
\end{figure}
We extend the analysis from the top five nodes to the entirety of FVS subsets by employing the area under the receiver operating characteristic curve (AUROC). The AUROC is a normalized score between 0 and 1 that measures how well a predictor metric (in this case, the propagation intersection metric) can sort a binarized metric (successful \textit{To/Away Control}) (see Methods). A value of 0 represents sorting completely incorrectly, a value of 1 represents sorting completely correctly, and a value of 0.5 represents sorting equivalent to random choice. Figure \ref{fig:fig4} summarizes the AUROC values of each network for all 100 rule-sets. The AUROC is greater than 0.5 (denoted by the dashed line) for the majority of the rule-sets, with $\geq$ 89.5\% of rule-sets having an AUROC greater than 0.5 in both the one-node, and two-node FVS subset cases for both \textit{To Control} and \textit{Away Control}. This indicates that the propagation intersection metric is able to discern FVS subsets with good control for each rule-set. Even though the AUROC values are below 0.5 for some rule-sets, this is not unexpected; we have previously shown that the topological metrics do not fully determine the possible dynamical behaviors of a model and that the top ranking FVS subsets have a much-better-than-random predictive power \cite{Newby2022-Chaos}.

\subsubsection{Propagation metrics accurately predict the logical domain of influence (LDOI) of a node}

The LDOI of a given node state consists of all the nodes whose value becomes fixed as the node state is percolated through the Boolean functions (see Methods). While it does not directly measure a node's effect on the attractor repertoire, the LDOI presents a detailed, dynamics-based understanding of how driving a node into a sustained state affects the network, so it provides an excellent way to probe how a node's impact differs for each different rule-set. Here we determine the consistency between the structural PRINCE and modified PRINCE metrics and the dynamics-based LDOIs.

We assess the consistency between the propagation scores $\Pi_j(t)$ and the LDOI by finding how often a perturbed node contains each other node in its LDOI. For each rule-set, we find the LDOI of fixing the perturbed node $n_i$ into a specific state $s_i$. This LDOI consists of a set of node-state pairs indicating which nodes are fixed and what state they are fixed into $LDOI(n_i, s_i) = \{(n_j, s_j)\}$. For each rule-set $r$, we give every node in the network one of three values based on its consistency with the perturbed node's fixed state: $L^{(n_i,s_i)}_r(n_j) = 1$ if the node is driven to the same state as the perturbed node, i.e., if  $s_j = s_i$, $L^{(n_i,s_i)}_r(n_j) = -1$ if the node is driven to the opposite state of the perturbed node, i.e., if $s_j = 1-s_i$, and $L^{(n_i,s_i)}_r(n_j) = 0$ if the node is not driven by the perturbed node, i.e., $n_j \notin LDOI(n_i, s_i)$. We add each node's values over all 100 rule-sets to get a number  of rule-sets in which a node is driven in a consistent manner in response to the perturbed node, which we refer to as the LDOI score $L^{(n_i,s_i)}(n_j) = \sum_{r=1}^{100}L^{(n_i,s_i)}_r(n_j)$. We differentiate between driving into the same state or the opposite state to discern nodes that drive consistently in the same direction over multiple rule-sets. For example, a case in which fixing a node ON drives another node ON in every rule-set will have a higher value than a case in which fixing a node ON drives another node ON in some rule-sets and OFF in other rule-sets. This process is done separately for fixing the perturbed node OFF, $s_i = 0$ and ON, $s_i = 1$. Thus, for each perturbed node $n_i$, every node in the network has two different LDOI scores.

\begin{figure}[htpb]
    \centering
    \includegraphics[width=\linewidth]{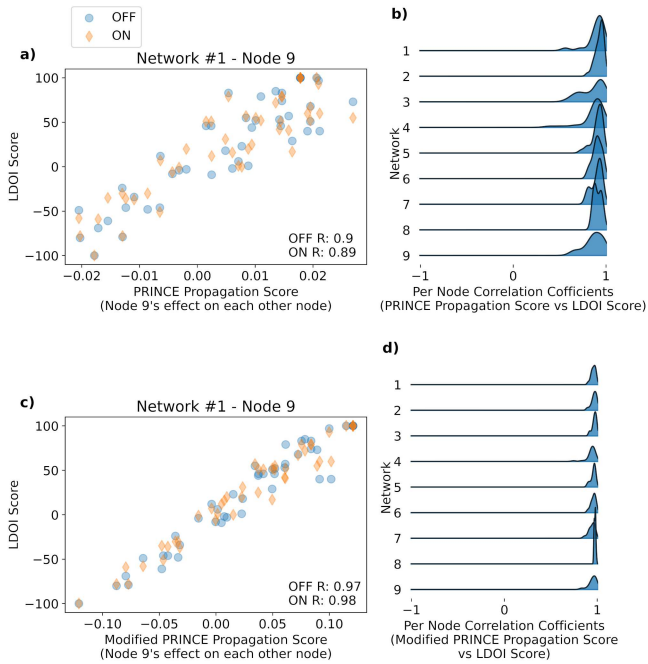}
    \caption{Comparison between a node's propagation score and its LDOI score. The propagation scores vs the LDOI scores for $s_i = 1$ (ON) and $s_i = 0$ (OFF) for PRINCE propagation (a, b) and modified PRINCE propagation (c, d) are shown. Panels a and c are sample plots of the propagation scores and LDOI score, using network \#1 and node 9 as the perturbed node. Panels c and d generalize these results using ridgeline plots of the PCC values for PRINCE propagation (b) and for modified PRINCE propagation (d) for every single-node FVS subset, including the PCCs with the LDOI scores for both ON and OFF perturbations.}
    \label{fig:fig5}
\end{figure}

Figure \ref{fig:fig5} shows the LDOI scores $L^{(n_i,s_i)}(n_j)$ vs the PRINCE scores $\Pi^{PRINCE}_j(t)$ (Fig. \ref{fig:fig5}a) and modified PRINCE scores $\Pi^{MPRINCE}_j(t)$ (Fig. \ref{fig:fig5}c) for each node $n_j$ in network \#1 when using node 9 as the perturbed node $n_i$. These plots reveal a strong correlation between both propagation metric scores and the LDOI scores: the Pearson correlation coefficients (PCCs) of the PRINCE score and LDOI score is 0.9 and 0.89 for fixing the perturbed node ON ($s_i = 1$) and OFF ($s_i = 0$) respectively with p-values of the order $10^{-17}$; the PCCs of the modified PRINCE score and LDOI score are 0.97 and 0.98 for ON and OFF respectively with p-values of the order $10^{-28}$. These results indicate that the propagation scores capture the ability of node 9 to drive the other nodes across the ensemble of models. To verify the generality of this conclusion, we determine the distribution of PCCs for both the ON state and OFF state of each FVS node in each network; see Fig. \ref{fig:fig5}b (PRINCE) and \ref{fig:fig5}d (modified PRINCE). The distribution of correlation coefficients shows that the strong relationship between propagation score and the LDOI score holds for all FVS nodes in every network topology.


\begin{figure*}[htpb]
    \centering
    \includegraphics[width=\textwidth]{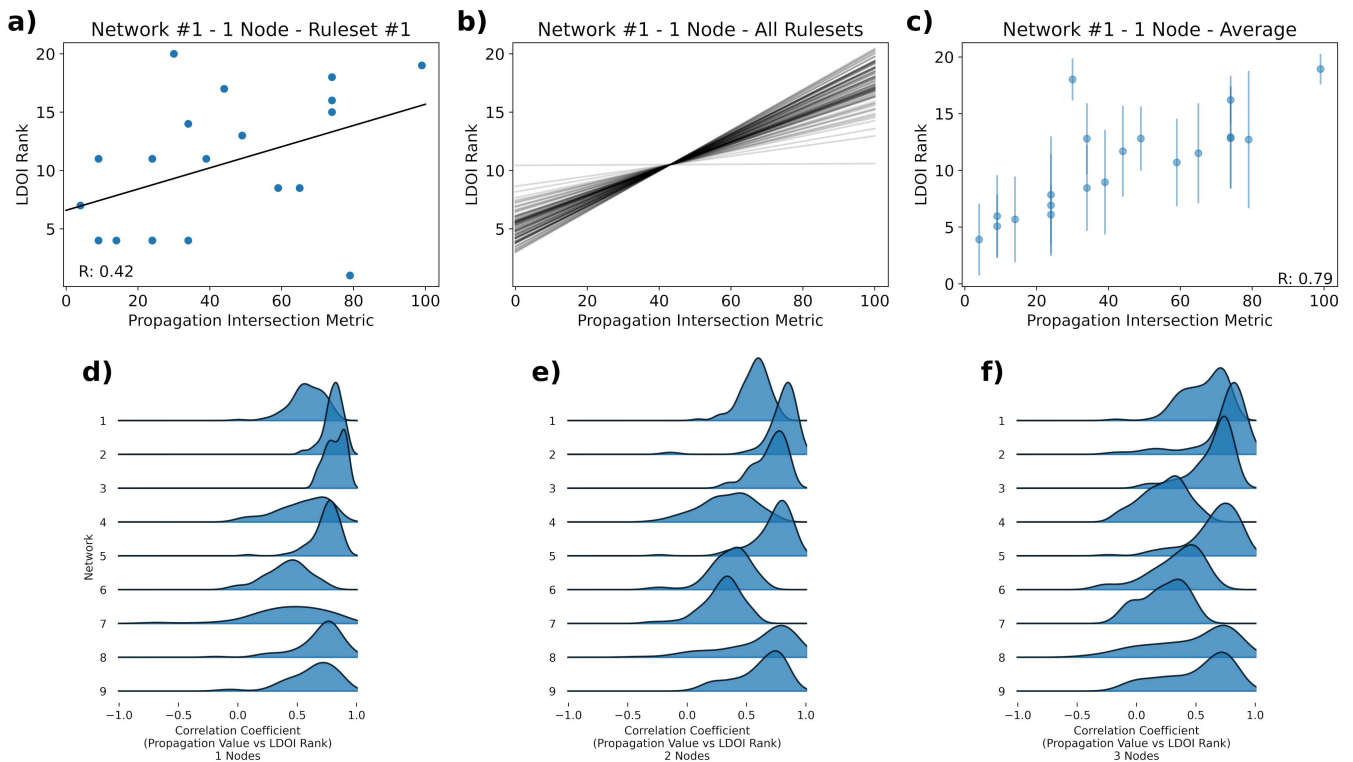}
    \caption{Scatter plot of each one-node FVS subset's propagation intersection metric value versus the rank of the FVS subset's LDOI size for the first rule-set generated on network \#1 along with the associated linear regression (a). The PCC for this data is printed in the bottom left corner. The linear regressions of all 100 of network \#1's rule-sets (b). Scatter plot of the propagation intersection metric versus average LDOI rank (c). The PCC for this data is printed in the bottom right corner. Distributions of the 100 rule-set's PCCs between the propagation intersection metric and LDOI rank of the one-node FVS subsets (d), two-node FVS subsets (e), and three-node FVS subsets (f).}
    \label{fig:fig6}
\end{figure*}

For each topology, we also found the propagation intersection metric value of every FVS node and compared these values to the size of the FVS node's LDOI in each individual rule-set. Specifically, for each FVS node, we calculate the size of the LDOI when fixing the node OFF and ON and take the maximum of these two values to give one LDOI value per node $\max{(|LDOI(n_i, 0)|,|LDOI(n_i, 1)|)}$. We then compare the ranks of the LDOI values with the propagation intersection metric values.  When ranking the subsets by LDOI size, the subset with the largest LDOI value gets the highest rank and the subset with the smallest LDOI value gets a rank of 1. Figure \ref{fig:fig6}a shows the propagation intersection metric values versus the LDOI ranks for each FVS subset of one rule-set of network \#1. The two values are correlated, which can be summarized with a linear regression. Figure \ref{fig:fig6}b shows the linear regressions of all 100 rule-sets of network \#1. All linear regressions except one have a positive slope, indicating that for 99\% of network \#1's rule-sets there is a positive correlation between the propagation intersection metric and the size of the LDOI for FVS subsets.

To test whether the observed correlation between propagation intersection metric values and LDOI ranks extends to all network topologies, we calculated the Pearson correlation coefficient of these metrics for each rule-set and each network topology. Figure \ref{fig:fig6}d-e shows the distributions of  PCCs over all 100 rule-sets for all nine network topologies. Similar to what we found for network \#1, the correlation coefficient is greater than zero for almost all rule-sets in each network topology; in the one-/two-/three-node case, 98\%/97\%/93\% of rule-sets have positive correlations. These positive correlations support the generality of the positive relationship between the LDOI and the propagation intersection metric. 

To further quantify the relationship between these metrics for each network topology, we calculated the Pearson correlation coefficient between the propagation intersection metric and the average LDOI rank over all 100 rule-sets. Fig. \ref{fig:fig6}c shows the results for one-node FVS subsets in network \#1, and table \ref{tab:table1} shows the correlation coefficient for FVS subsets sizes 1 to 3 for all nine network topologies. Of the nine network topologies, the smallest correlation coefficient was 0.53 and the largest was 0.97.

\begin{table}[ht]
\caption{\label{tab:table1}
Pearson correlation coefficients (PCCs) of the average LDOI rank and the propagation intersection metric values. These are the correlations between each FVS subset's average LDOI rank over all 100 rule-sets and the propagation intersection metric value. We present the results for all nine networks for FVS subsets of sizes 1-3.
}
\begin{ruledtabular}
\begin{tabular}{lccccccccc}
Network & 1 & 2 & 3 & 4 & 5 & 6 & 7 & 8 & 9\\
\hline
1 Node & 0.79 & 0.97 & 0.89 & 0.86 & 0.96 & 0.74 & 0.71 & 0.96 & 0.91\\
2 Node & 0.76 & 0.95 & 0.86 & 0.66 & 0.95 & 0.65 & 0.60 & 0.94 & 0.91\\
3 Node & 0.80 & 0.90 & 0.88 & 0.53 & 0.93 & 0.93 & 0.59 & 0.88 & 0.87\\
\end{tabular}
\end{ruledtabular}
\end{table}

Overall, the results of our analysis of the successfulnesses, AUROCs, and correlation coefficients all indicate that there is a strong relationship between the structural propagation metrics and the controllability of the system reflected by the control metrics of \textit{To Control, Away Control}, and LDOI. These results are consistent for both the averages of all 100 rule-sets and on an individual rule-set level, showcasing that the FVS subsets found using the propagation intersection metric can typically control a system without consideration for the specifics of the dynamical model.

\subsection{Propagation metrics on the effective graph}
Despite the generally good  accuracy of the structure-based propagation metrics in predicting controllability through the \textit{To Control, Away Control}, and the LDOI, they do not fully capture the dynamical behaviors of the system. A likely reason for this shortcoming is that propagation metrics solely take into account structure and are not dynamic model-specific. A way to improve the predictive power of the propagation metrics would be to incorporate information of the dynamical model into the network topology. One way to do this is the effective graph \cite{Gates2021-PNAS}. The effective graph assigns each edge in the network a weight based on that edge's “effectiveness”, which measures how important that edge is in determining the output of the associated Boolean function (see Methods). Thus, the resulting network encodes features of the regulatory functions into its structure.

Given that the effective graph of a Boolean model is a weighted network, we can calculate the propagation metrics on the effective graph. The value of the propagation metrics in the effective graph captures how information propagates through the network, biasing information transfer through the edges based on their associated Boolean functions. This means that the nodes with the highest metric values are more likely to fix the outcome of the Boolean functions in addition to being structurally important. We compare the propagation metrics calculated on each Boolean rule-set's associated effective graph to the original network's propagation metrics. Using this comparison, we assess the impact of including functional information within the network's structure to gauge how much the propagation metrics are gated by being only topological.

\subsubsection{The effective graph's propagation metrics are consistent with those of the original graph}

\begin{figure}[H]
    \centering
    \includegraphics[width=\linewidth]{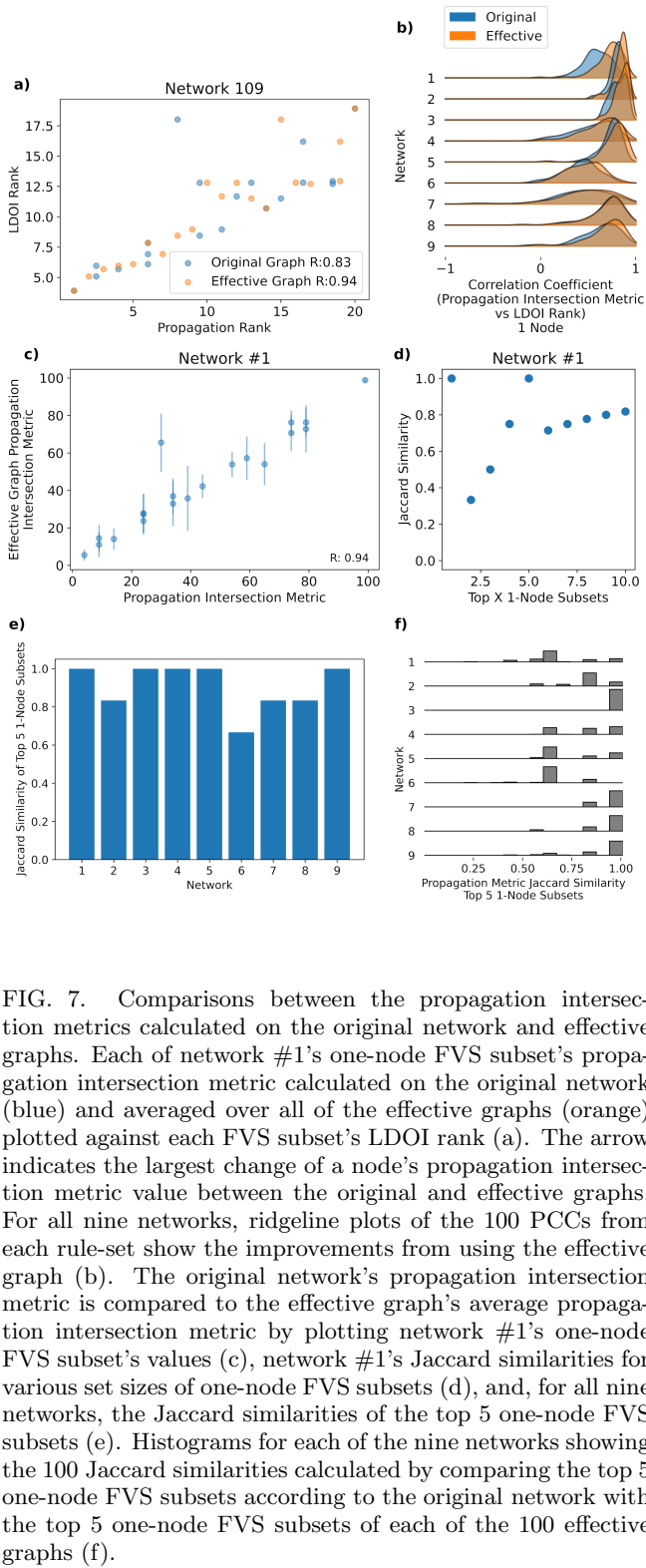}
    \caption{Comparisons between the propagation intersection metrics calculated on the original network and effective graphs. Each of network \#1's one-node FVS subset's propagation intersection metric calculated on the original network (blue) and averaged over all of the effective graphs (orange) plotted against each FVS subset's LDOI rank (a). The arrow indicates the largest change of a node's propagation intersection metric value between the original and effective graphs. For all nine networks, ridgeline plots of the 100 PCCs from each rule-set show the improvements from using the effective graph (b). The original network's propagation intersection metric is compared to the effective graph's average propagation intersection metric by plotting network \#1's one-node FVS subset's values (c), network \#1's Jaccard similarities for various set sizes of one-node FVS subsets (d), and, for all nine networks, the Jaccard similarities of the top 5 one-node FVS subsets (e). Histograms for each of the nine networks showing the 100 Jaccard similarities calculated by comparing the top 5 one-node FVS subsets according to the original network with the top 5 one-node FVS subsets of each of the 100 effective graphs (f).}
    \label{fig:fig7}
\end{figure}

We determine the potential of the effective graph to identify drivers of the network by comparing the effective graph propagation intersection metric values with the original network propagation intersection metric values. Figure \ref{fig:fig7}a showcases an example of the improvements that can be gained by using the effective network to calculate the propagation metrics. Here, we plot the propagation intersection metric of the original network (blue) and the average propagation intersection metric of the 100 effective graphs (orange) vs the LDOI rank of each one-node FVS subset for network \#1. For the original network, we get an adequate PCC of 0.79, but when using the averaged effective graphs, this correlation is improved to 0.92.

The main reason for this improvement is the increase of node 48's propagation intersection metric. In the original network, node 48 is an outlier with a low propagation intersection metric value and a high LDOI rank. The reason for the originally low propagation metric is a negative self-loop on the node. Yet, node 48 has four other inputs, so, in a majority of rule-sets, node 48's self-loop does not cause oscillations. The relative unimportance of the negative self-loop is captured by the effective graphs, allowing node 48's propagation intersection metric value to be larger in the effective graphs (indicated by the arrow in Fig. \ref{fig:fig7}a). These types of changes improve the accuracy of the propagation intersection metric in the effective graphs because less impactful edges impact the propagation intersection metric less.

We also investigated how much the correlation between the propagation intersection metric and the LDOI improved for individual rule-sets (Fig. \ref{fig:fig7}b). Here, we calculated each FVS subset's propagation intersection metric on the original network, and, for each rule-set, we calculated each FVS subset's effective graph propagation intersection metric. We found the correlation of these values with the LDOI ranks of every one-node FVS subset. For each network, we calculated the 100 PCCs of the propagation intersection metric or effective graph propagation intersection metric vs the LDOI ranks and plotted these correlations. The effective graph provides an improvement to correlation coefficients across all of the rule-sets, but the improvement is typically not drastic. We conclude that, while incorporating the Boolean functions into the effective graph brings a tangible improvement, the structure alone still leads to reasonably successful results when lacking functional information.

We next directly compare the FVS subsets's propagation intersection metric in the original network and in the effective graph. We find the propagation intersection metric values of every FVS subset in both cases and explore the relationship between the values. We first compared the values of the FVS subset's propagation intersection metrics calculated on the 100 weighted effective networks, and found that they are very similar to those calculated on the original, unweighted network. Figure \ref{fig:fig7}c shows the propagation intersection metric and the average propagation intersection metric of the effective graphs for network \#1. The values of the propagation intersection metrics between the original and the effective graph are highly concordant and well correlated (PCC = 0.94). Looking at each network we also find the intersection metrics to be highly concordant (PCC range of 0.91-1.0, see Table \ref{tab:table2}).

\begin{table}[ht]
\caption{\label{tab:table2}
Pearson correlation coefficients of the propagation intersection metric calculated on the original network vs the average propagation intersection metric calculated on the 100 effective graphs for each of the nine networks.
}
\begin{ruledtabular}
\begin{tabular}{lccccccccc}
Network & 1 & 2 & 3 & 4 & 5 & 6 & 7 & 8 & 9\\
\hline
PCC & 0.94 & 1.0 & 0.97 & 0.99 & 0.99 & 0.91 & 0.99 & 0.99 & 0.99\\
\end{tabular}
\end{ruledtabular}
\end{table}

To focus on the possible control subsets identified by our approach, we compare the sets of top ranking FVS subsets. We quantify the similarity between these sets in the original network and the effective graph using the Jaccard similarity. The Jaccard similarity is defined as $(S_1 \cap S_2)/(S_1 \cup S_2)$ where $S_1$ is the group of top FVS subsets according to the original propagation intersection metric and $S_2$ is group the top FVS subsets according to the average effective graph propagation intersection metric. Figure \ref{fig:fig7}d shows the Jaccard similarities for the one-node FVS subset case of network \#1. The top 5 FVS nodes are the same in both the original and effective graphs, showing that the most dynamically relevant nodes in the effective graph are also identified in the unweighted, original case.

Looking at each of the nine network topologies, we found that there are high Jaccard similarities between the propagation intersection metric of the original network and average propagation intersection metric of the effective graphs. Figure \ref{fig:fig7}e shows the Jaccard similarities of the top 5 one-node FVS subsets for all nine networks. For the top 5 FVS subsets, the lowest similarity was that of network \#6, and has a value of 0.67 (i.e., the top 5 FVS subsets of both sets share 4 subsets). For network \#2 and network \#7, the Jaccard similarity is 0.83 (i.e., the two sets share 5 subsets, but one of the sets is of size 6 because of a tie). Figure \ref{fig:fig7}f shows histograms of the 100 Jaccard similarities calculated when comparing the top 5 FVS subsets of the original network and each rule-set's effective graph. A vast majority of the rule-sets achieve a similarity above 0.5, indicating that more than 3 out of 5 of the identified subsets were in both sets. 

In summary, we find that the propagation metrics when calculated on a weighted version of the network where weights approximate functional value do more accurately capture the dynamics than the propagation metrics calculated on the original network. Therefore, the accuracy of the propagation metrics is limited by only having knowledge of the system's structure. We further show that while the effective graph provides an improvement to the propagation metrics' ability to capture the Boolean model, the benefits are not substantial. Many of the FVS subsets that are improved by utilizing the effective graph are not the top ranking FVS subsets. The top performing propagation metrics calculated on the original network and on the weighted effective graphs are highly concordant as shown by the Jaccard similarities of the two sets, so the FVS subsets that are most likely to drive the network according to the propagation intersection metric are the same between the effective network and the original network. These results further reinforce our main findings throughout the manuscript: that the top performing FVS nodes according to the propagation intersection metric are good at controlling the network and are robust to the details of the dynamical model.

\section{Discussion}

\begin{figure*}[htpb]
    \centering
    \includegraphics[width=\textwidth]{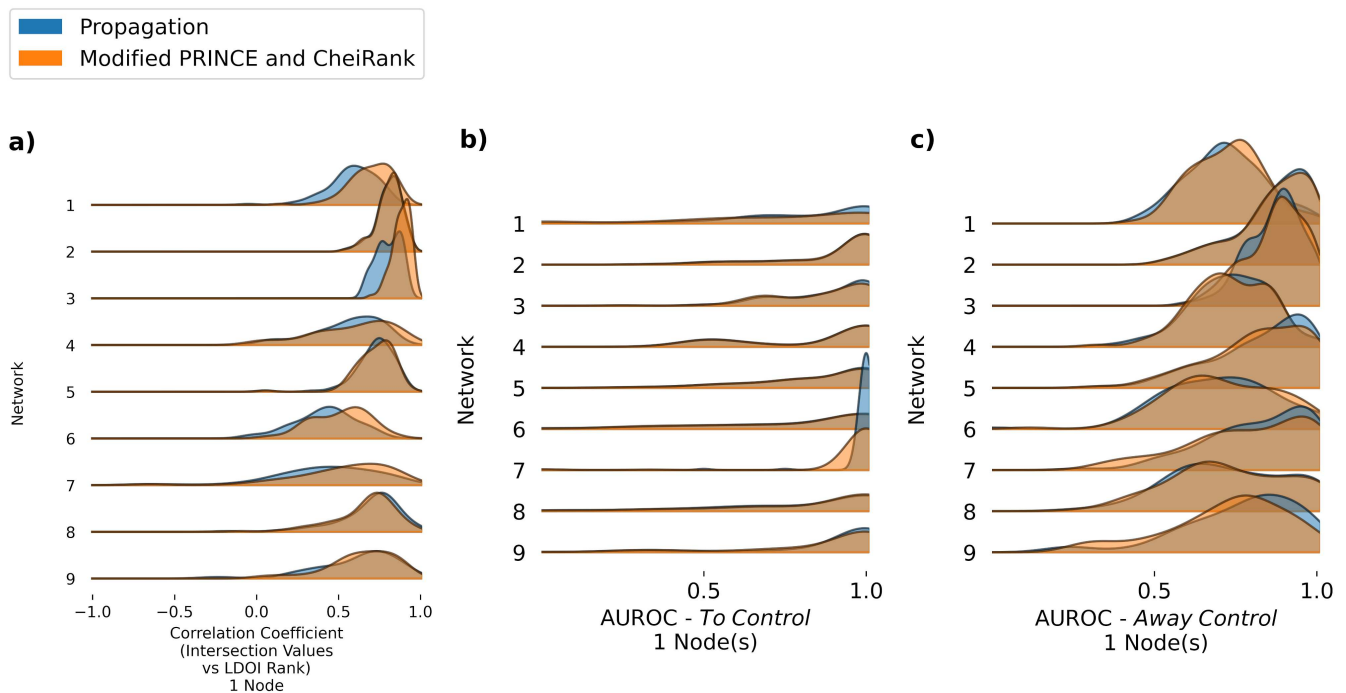}
    \caption{Comparisons between two intersection metrics of the propagation metrics: the intersection of all three propagation metrics (blue) and the intersection of just the modified PRINCE and CheiRank metrics (orange). The distributions of the Pearson correlation coefficients between intersection values and LDOI rank (a) and the AUROCs of \textit{To Control} (b) and \textit{Away Control} (c) for all nine networks.}
    \label{fig:fig8}
\end{figure*}

Ensembles of random Boolean networks have a long track record of being used to reveal the emergent properties of biological systems. As our understanding of biological systems improves, the randomly generated networks that we develop to reflect these systems improve too. The network ensembles used here, reflect both the structure and function of biological systems by using power law out-degree, Poisson in-degree distributions and nested canalyzing functions. Using this information, we generate ten networks and an ensemble of 100 Boolean rule-sets for each network. This provides us with a large test bed of networks to thoroughly probe how well structurally significant node sets can control a network. On these ensembles of Boolean models, we investigated whether structural propagation metrics can identify attractor control sets that are robust to the specifics of the Boolean model.

In attractor control theory, the feedback vertex set guarantees a system will be driven to a target attractor, so the subsets of the FVS are prime candidates for driving the network to or away from attractors. The FVS provides information about the system's cycle structure and thus, its attractor repertoire. As a result, FVS subsets are predisposed to having some control over a system's attractors \cite{Newby2022-Chaos}. The propagation metrics — PRINCE propagation, modified PRINCE propagation, and CheiRank — are able to utilize both cycle and path structures in the network to rank FVS subsets by their likelihood of controlling the network's trajectory. Here, we have shown through \textit{To Control, Away Control}, LDOI, and effective graphs that FVS subsets identified using the propagation metrics can successfully drive networks towards or away from its attractors.

When looking at the top-ranking FVS subsets according to the propagation intersection metric, we find that the identified FVS subsets successfully control most rule-sets according to both \textit{To Control} and \textit{Away Control}. Furthermore, when analyzing individual rule-sets, the AUROCs of most rule-sets are greater than 0.5 further indicating that the propagation intersection metric is predictive of attractor control. While \textit{To Control} and \textit{Away Control} are better at measuring attractor controllability, the LDOI provides a more detailed view of each node's influence on the network. The LDOI score is highly consistent with the PRINCE and modified PRINCE scores signaling that these two propagation metrics reflect the average reach of an FVS node. Furthermore, the propagation intersection metric accurately predicts a FVS subset's LDOI rank, as shown by the Pearson correlation coefficients between these measures. This indicates that the propagation metrics are able to approximate the Boolean dynamics of the network because the propagation metrics are able to predict the simpler Boolean functions, which make up a majority of the network's interactions. The propagation metrics do not perfectly reproduce the dynamics because they cannot capture the more complicated combinatorial nature of Boolean functions. The ability for the propagation metrics to capture biological functions may be related to the recent result that biological functions tend to be less non-linear than expected, which is better reflected in nested canalyzing functions than random Boolean functions \cite{Manicka2023-bioRxiv}. Thus, while the propagation metrics make simplifying assumptions about the combinatorial nature of Boolean functions, because a majority of the functions are simple, they do well approximate the dynamics of the system.

\begin{figure*}[htpb]
    \centering
    \includegraphics[width=\textwidth]{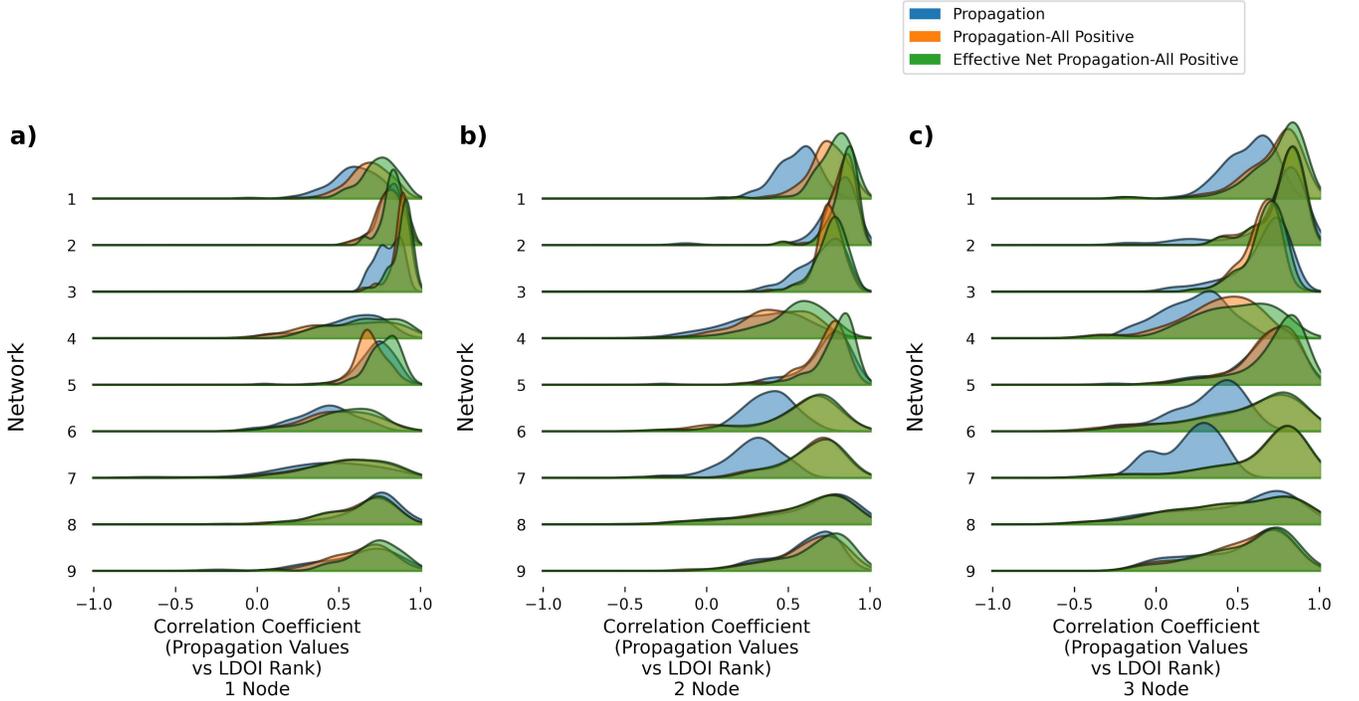}
    \caption{Comparison of propagation intersection metric calculated on modified versions of the nine networks. The propagation intersection metric calculated on the unweighted original network (blue) is compared to the propagation intersection metric calculated on a modified network with only positive edges (orange) and on the weighted effective graphs of the networks with only positive edges (green). Distributions of the nine network's Pearson correlation coefficients between the propagation values and LDOI ranks for one- (a), two- (b), and three-node (c) FVS subsets for each of the different network variations.}
    \label{fig:fig9}
\end{figure*}

When comparing the distributions of correlation coefficients in Fig. \ref{fig:fig5}b and Fig. \ref{fig:fig5}d it is clear that the LDOI is more strongly correlated to the modified PRINCE propagation than the PRINCE propagation. This is unsurprising because the modified version of PRINCE propagation was originally developed to reflect the propagation of information, which does not obey conservation rules akin to mass conservation \cite{Newby2022-Chaos}. Information propagation and regulation unrestricted by mass conservation is more common in the biological networks modeled by Boolean models. We investigated if there would be a considerable benefit to only using the modified PRINCE propagation and CheiRank instead of all three propagation metrics, but found that there were no ubiquitous improvements in our analyses (Fig. \ref{fig:fig8}). While using only the modified PRINCE and CheiRank intersection metric does boost the PCC between intersection metric value and LDOI in some cases and never decreases the PCCs (Fig. \ref{fig:fig8}a), \textit{To Control} and \textit{Away Control} AUROCs are little affected by the change of metrics (Fig. \ref{fig:fig8}b,c). Because we are focused on attractor control and the predictive power with respect to \textit{To Control} and \textit{Away Control} shows little difference, using all three propagation metrics for our analysis is a reasonable choice.

The effective graph, which incorporates functional information into the network structure, can be used to explore the structural limitations of the propagation metrics. The propagation metrics when calculated on the effective graphs showcase an improvement of the correlation with the LDOI rank. However, when focusing on the top performing FVS subsets, there is high similarity between the original network and each effective graph, so the top performing FVS subsets are the top performing subsets in every rule-set. Employing the effective graph provides a way to improve the overall correlation between propagation intersection metric and LDOI, but it does not improve the identification of the top attractor control sets. Instead, the effective graph helps to better classify structural outliers (see node 48 in Fig. \ref{fig:fig7}a). Thus, if available the propagation intersection metric on the effective graph is a more accurate representation, but if only structural information is available, the propagation intersection metric on the original network is suitable for identifying driver sets.

We have shown that the propagation intersection metric can be used to rank FVS subsets on both biological networks \cite{Newby2022-Chaos} and on biologically-inspired networks. In these types of networks, the structure of the network indicates nodes that are predisposed to driving the network to its attractors. While it is not clear if this approach generalizes to arbitrary network structures, it works accurately for networks that have a structure and function similar to typical of Boolean biological networks (i.e., it has a power law out-degree distribution, a Poisson in-degree distribution, and is defined by nested canalyzing functions). This approach does have two limitations. First, the propagation metrics only capture the network's structure, but we have shown through the effective graphs that the improvements gained from incorporating features of the regulatory functions into the propagation metrics does not largely affect the top identified drivers. The second issue is that while the PRINCE and modified PRINCE propagations do incorporate negative edges into their values, none of the propagation metrics can accurately describe the effects that negative feedback loops have on the network's dynamics. These limitations can be seen in Fig. \ref{fig:fig9} where on a modified version of the networks with only positive edges every network achieves higher correlation coefficients. Further modifying the networks using the effective graphs to incorporate functional information further improves the correlation coefficients. More work is needed to accurately capture the impact of negative feedback loops on the dynamics using only the structure, so we expect that the propagation intersection metric will be less accurate for biological networks consisting of many complex attractors. Despite these limitations, on a typical biological network we expect the propagation metrics to accurately identify attractor controlling FVS subsets.

Because the FVS and propagation metrics are structural measures, knowledge of a biological network's dynamical model is not necessary to extract control information. When constructing biological networks, the specifics of the update functions are typically formulated based on incomplete information, so in some cases only the existence of a connection between nodes is known and in other cases the update function is only the best available approximation of the biological system's behavior. Since our approach does not utilize any dynamical information, it can be used both as a prediction tool (i.e., what are the top driver nodes) and a validation tool (i.e., are the top driver nodes consistent with current knowledge) on less well defined models. In networks without a dynamical model, this approach can be used to identify key nodes to further investigate instead of needing detailed knowledge about every interaction to build the dynamical model first before finding key nodes. As previously mentioned, this will not find every possible key node, which cannot be done without the full dynamical model. Furthermore, as the FVS was originally defined for networks with ODE update functions, this approach should be applicable as a prediction tool in other dynamical models than Boolean models. In its original formulation, the FVS was defined for dissipative systems (i.e., a system where variables cannot increase infinitely), which biological systems typically are. The propagation metrics ability to capture a system's dynamical evolution has not been exhaustively proven for ODE systems, but it has been demonstrated on continuous systems \cite{Santolini2018-PNAS}. Because both the FVS and propagation metrics are likely extendable to both non-Boolean discrete systems and continuous systems, this approach should be viable for making control predictions on these systems as well. As a validation tool, in networks with a dynamical model, the node subsets identified by this approach can be compared to dynamically identified node subsets. If these dynamically important nodes are also structurally important, then the dynamical model is consistent with an “average” biological network's model. If these sets are not similar, caution must be taken to verify that this is not due to overfitting or low-confidence parameters in the dynamical model.

In summary, using the propagation metrics we can identify subsets of the FVS that are able to control a biological network regardless of the specifics of the Boolean model. Here, using biologically inspired network structures with ensembles of Boolean models composed of nested canalyzing functions, we have extensively shown the ability of propagation metrics to predict key attractor driving sets. These sets are, by construction, subsets of the network's FVS, which is composed of the nodes with the largest influence over the network's multistability. Propagation metrics are able to rank FVS subsets by approximating the Boolean dynamics of the simpler Boolean functions that make up a majority of interactions in a biological network model. Thus, utilizing the top-ranking FVS subsets according to the propagation intersection metric, we are able to drive a typical biological network towards or away from a target attractor without needing detailed knowledge of the network's dynamical model.

\begin{acknowledgments}
This work was supported by Army Research Office grant 79961-MS-MUR and National Science Foundation grant MCB 1715826. We thank Dr. Jordan Rozum, Fatemeh Sadat Fatemi Nasrollahi, and Kyu Hyong Park for helpful discussions.
\end{acknowledgments}

\section*{Author Declarations}
\subsection*{Author Contributions}
E.N conceptualized the idea and J.G.T.Z. and E.N. developed the methodology with input from R.A.. R.A. acquired funding and provided project administration. E.N. programmed the software with help from J.G.T.Z.. E.N. investigated the problem through conducting simulations, and provided formal analysis of the resulting data under the supervision of both R.A. and J.G.T.Z.. E.N. visualized the data and wrote the original draft, and J.G.T.Z. and R.A. reviewed and edited the final work. J.G.T.Z. and R.A. contributed equally to this work as senior authors.
\subsection*{Conflicts of Interest}
The authors have no conflicts to disclose.
\subsection*{Data Availability}
The data that support the findings of this study are available from the corresponding author upon reasonable request.

\appendix
\section{Methods\label{append:Methods}}
\subsection{Propagation Intersection Metric}

The propagation intersection metric \cite{Newby2022-Chaos} is calculated by combining the three propagation metrics: PRINCE propagation \cite{Santolini2018-PNAS}, modified PRINCE propagation \cite{Newby2022-Chaos}, and CheiRank \cite{Zhirov2010-EPJB}. These three metrics calculate how information is propagated through the network. The two PRINCE propagation metrics calculate an information score for every node based on starting a constant perturbation at an initial subset of nodes and measuring how that information diffuses throughout the system. Both metrics will reach a steady state showing the perturbed subset's effect on the rest of the network. We average the resulting information scores of each node to determine the metric value for the perturbed subset. The CheiRank is an extension of the PageRank that is calculated in the same manner as the PageRank but by traversing the edges in reverse. It is a calculation of the “source-ness” of a node. That is, it is a calculation of the probability that a random walker began walking at any given node, which indicates that it is more likely to be integral in driving the network.

After calculating the values of every FVS subset, we calculate each FVS subset's percentile rank for all three metrics. For each metric, we find the set of FVS subsets with percentile rank higher than a given percentile value. We find the highest percentile value that includes a FVS subset in all three sets, which combines the three percentile ranks into a single value. This highest percentile value is the propagation intersection metric value for each FVS subset. This is equivalent to finding the minimum percentile rank for each FVS subset among the three propagation metrics.

\subsection{\textit{To Control} and \textit{Away Control}}

When simulating a network under an intervention, we measure the intervention's success through \textit{To Control} and \textit{Away Control} \cite{Newby2022-Chaos}. These two metrics measure how the intervention has affected the system's basins of attraction. \textit{To Control} measures how much the target attractors' basins increases, and \textit{Away Control} measures how much the non-target attractors' basins decreases. To calculate these values, we first simulate the unperturbed network 100 times to approximate the size of the wild-type basins of attraction. Next for each FVS subset of size \textit{L}, we intervene by fixing all $2^L$ possible input combinations of the set $\{0,1\}^L$. For each input combination, we determine if the intervention is \textit{not informative}, \textit{partially informative}, or \textit{fully informative}. The first two cases are due to the nodes in the FVS subset having the same state in all the wild-type attractors. A \textit{not informative} intervention targets every wild-type attractor; the intervention cannot increase the basins of the targets and there are no non-targets. A \textit{partially informative} intervention targets none of the wild-type attractors; the intervention can decrease the basis of non-targets, but there are no targets, so we can measure \textit{Away Control}, but we cannot measure \textit{To Control}. A \textit{fully informative} intervention targets some of the wild-type attractors, so both \textit{To Control} and \textit{Away Control} can be calculated. In the relevant cases, \textit{To Control} is calculated by the percent change in the sum of the target attractors' basins when the intervention is applied compared to the wild-type basins. Similarly, \textit{Away Control} is calculated by the percent change in the sum of the non-target attractor's basins.

\subsection{AUROC}

A receiver operating characteristic (ROC) curve plots the tradeoff between the false positive rate and true positive rate when looking at various sorting thresholds of a binary classifier on a data set. In our case the propagation intersection metric sorts the binarized successful \textit{To Control} or successful \textit{Away Control}. The false positive rate (FPR) is the ratio of the number of false positives to the total number of negative cases. The true positive rate (TPR) is the ratio of the number of true positives to the total number of positive cases. To plot the ROC, the binary classifier is scanned over all values, and the FPR and TPR are calculated at each value. The FPR and TPR are then plotted against each other resulting in a curve that demonstrates how well the binary classifier classifies the data over all threshold values.

The area under the ROC curve (AUROC) quantifies the classification strength of the binary classifier. Because the FPR and TPR are both ratios, the AUROC is normalized between 0 and 1. A value of 1 indicates that the binary classifier perfectly sorts the data (i.e., the high propagation intersection metric FVS subsets have high \textit{To/Away Control}). A value of 0 indicates that the binary classifier sorts completely incorrectly (i.e., high propagation interaction metric FVS subsets have low \textit{To/Away Control} values). An AUROC of 0.5 indicates that the classifier functions equivalently to randomly choosing if the data point has high or low \textit{To/Away Control}.

\subsection{LDOI}
The logical domain of influence (LDOI) identifies the nodes that directly or indirectly become fixed in a node state when a node is fixed to a specific state \cite{Yang2018-FrontPhysio}. Each node has two LDOI values, one for each of its possible states. For each state, the LDOI is calculated by a percolation process, i.e., by identifying the nodes whose Boolean function stabilizes to a single state when the initial node state is fixed, and iteratively repeating this procedure for each node  whose Boolean function is stabilized. If this procedure reaches a contradiction (e.g., a node state opposing the original node state that would be stabilized), the percolation is stopped  because it is assumed that the original node state will be sustained. When consolidating the LDOI to a single value per node, we take the maximum between the LDOIs of the ON and OFF states. The LDOI can also be found for fixing a set of nodes instead of a single node.

\subsection{Effective Graph}

The effective graph is a directed weighted network wherein the edge weights represent the effectiveness of that edge in fixing the output of the Boolean function of the target node \cite{Gates2021-PNAS}. To quantify the effectiveness of each variable in a Boolean function, we first calculate their redundancy. The redundancy of an input variable is the fraction of input combinations where the variable does not influence the output (i.e., the output is the same for both values of the input variable). The effectiveness is the complement to the redundancy (i.e., effectiveness is $1-redundancy$). Once the effectiveness is calculated for every input variable for every function, we can use it as a weight of the edge from the input variable to the target node and generate the effective graph. 

The effective graph provides a way to append dynamical information to the topological structure of the network through edge weights. We then evaluate the propagation metrics using the weighted adjacency matrix associated with the effective graph. This gives a more accurate representation of how information is propagated through the network because the propagation is weighted by the dynamical model. We compare these propagation metrics from each individual rule-set to the metrics calculated on the original topology to determine how dependent the control of the system is on the dynamical rules.

\subsection{\label{subsec:Computational}Computational implementation}

These methods were implemented in python using various libraries and modules. The code for identifying, ranking, and calculating \textit{To Control} and \textit{Away Control} of FVS subsets is available at \url{https://github.com/EliNewby/FVSSubsets}. We include a Jupyter notebook example that shows the output of this code on a T-LGL network model. To identify near-minimal FVSs and their subsets, a python code developed by Gang Yang was used, \cite{Zanudo2017-PNAS} which utilizes the simulated annealing algorithm presented by Galinier et al. \cite{Galinier2013-Heur} This code is available at \url{https://github.com/jgtz/FVS_python3}. The simulations used to calculate \textit{To/Away Control} values were implemented using the bioLQM toolkit developed by the CoLoMoTo Consortium \cite{Naldi2018-Front}. Using bioLQM, we implement our Boolean models of the networks, find their attractors, and simulate the system's trajectories using the random asynchronous update mode. To calculate the LDOIs of each Boolean rule-set, we use the python library pyStableMotifs developed by Jordan Rozum \cite{Rozum2021-Bioinfo}. To generate each rule-sets effective graph, we use the python library CANA developed by Correia and Gates \cite{Correia2018-Front}.

\section{Formalization of feedback vertex set control\label{append:FVS}}
The concept of FVS control, proposed by Mochizuki, Fiedler and collaborators \cite{Fiedler2013-DynamDiffEQ, Mochizuki2013-JTheorBio}, applies to systems governed by a general class of nonlinear dynamics, for any network structures, and for any value of the parameters describing the dynamics. This theory shows that even in the absence of  detailed knowledge of the network's dynamics it is possible to drive the network into one of its natural attractors. A feedback vertex set (FVS) is a set of nodes that when removed from a network leaves an acyclic network. The essence of this theory is as follows: because the cycles of a network determine its attractor repertoire, a FVS has control over the attractor repertoire of the system. In the following we give the key details of the theory.

FVS control considers networks where the system's nodes states $x_i(t)$ are described by differential equations of the form \begin{equation}
\label{EqA1}
    dx_i/dt = f_i(x_{I_i})-d_i(x_i),\quad i = 1, 2,\ldots,N,
\end{equation}

 where $I_i$ is the set of regulators of node $n_i$. These equations include a decay term $d_i(x_i)$ that depends only on the current state of the node. The parameterization of the decay term ensures that the system is dissipative, which guarantees that the values of the variables do not infinitely increase and allows for the convergence of the system to attractors. When a node $i$ is self-regulated in such as way that $\partial{f_i}/\partial{x_i} > \partial{d_i}/\partial{x_i}$, node $i$ is interpreted as $i \subset I_i$ and a positive self-loop is added to $i$, making it part of the FVS.

Consider a system that satisfies the differential equations \ref{EqA1}, a given FVS denoted $V$, and two trajectories, $X(t)$ and $\widetilde{X}(t)$. If the two sub-trajectories, $X_V(t)$ and $\widetilde{X}_V(t)$ are related such that $X_V(t)-\widetilde{X}_V(t) \rightarrow 0$ in the limit $t \rightarrow \infty$, i.e., if the values of every FVS node in each trajectory's attractor are equal, then $X(t)-\widetilde{X}_(t) \rightarrow 0$ in the limit $t \rightarrow \infty$, i.e. both trajectories lead to the same attractor. In other words, an FVS is a set of "determining nodes"; fixing every node of a FVS to its respective state in a target attractor will fix the trajectories of every other node in the network and force the system into the target attractor. This conclusion applies for any value of the parameters involved in the functions in \ref{EqA1}.

FVS control functions by reducing the state space into a single dynamical attractor. Ren\'e Thomas demonstrated that multistability of a network is a consequence of its positive feedback loops \cite{Thomas1981-NumericalMethods, Thomas2001-Chaos, Thomas2001-Chaos2}, so an acyclic network will only have a single attractor. Thus, because removing a FVS removes every cycle from the system, it will also result in a system with only a single dynamical attractor. Likewise, driving the regulatory functions $f_{V} = (X_{V}, X_{I_{V}}, t)$ of every FVS node into its state in a desired attractor $\boldsymbol{X}_D$ (i.e., $f_{V} = f_{V}(\boldsymbol{X}_D)$) will reduce the multistability of the network and will result in the state space only containing one attractor, namely the desired attractor $\boldsymbol{X}_D$.

The originally defined framework \ref{EqA1} did not include the possibility of source nodes. These nodes are independent of the internal state of the system and can represent external inputs to the system. The framework was extended in \cite{Zanudo2017-PNAS} to include $N_s$ source nodes with variables $s_i(t)$
\begin{align}
    ds_i/dt &= g_i(t), \quad &i &= 1, 2, \ldots, N_s\\
    dx_j/dt &= f_j(x_{I_j}) - d_j(x_j), \quad &j &= N_s+1, \ldots, N
\end{align}
In these systems, for FVS control to drive the system into a desired attractor, the source nodes must be driven too. That is, if $s_i^{\boldsymbol{X}_D}(t)$ are the states of the source nodes in the desired dynamical attractor $\boldsymbol{X}_D$, then driving every source node such that $s_i(t) - s_i^{\boldsymbol{X}_D}(t) \rightarrow 0$ as $t \rightarrow \infty$ will reduce the system so that it can be defined by \ref{EqA1}. This reduced system can then be driven into the desired attractor following the same procedure as before of driving the nodes in a FVS to their given states in the desired attractor.

Boolean systems are not directly described by the framework of \ref{EqA1} because Boolean update functions show the state at the next time step ($f(t+dt)$) and not how the function changes in time ($df/dt$). Yet, Boolean systems align with the general concept: they are non-linear systems that are dissipative. Indeed, if at time-step $t-1$ node $i$ is activated and its regulators $I_i$ are in a state for which the node's Boolean function evaluates to 0, then the node will decrease to an inactive state at time $t$. In addition, a Boolean system's state space is restricted to an N-dimensional Euclidean hypercube, so every trajectory will always converge to an attractor. In summary, FVS control is also applicable to Boolean dynamical models.


\providecommand{\noopsort}[1]{}\providecommand{\singleletter}[1]{#1}%

\end{document}